# Low-power Rapid Planar Superconducting Logic Devices

Nikolay Gusarov,[1] Rajesh Mandal,[1,2] Issa Salameh,[3] Itamar Holzman,[1,2] Shahar Kvatinsky,[3] and Yachin Ivry[1,2,4,*]

**AFFLIATIONS**

[1]Department of Materials Science and Engineering, Technion - Israel Institute of Technology, Haifa 3200003, Israel

[2]Solid State Institute, Technion – Israel Institute of Technology, Haifa 3200003, Israel

[3]Andrew and Erna Viterbi Faculty of Electrical and Computer Engineering, Technion - Israel Institute of Technology, Haifa 3200003, Israel

[4]Department of Materials Science and Engineering, University of California, Berkeley, CA 94720, USA

**\*Author to whom correspondence should be addressed: ivry@technion.ac.il**

**ABSTRACT**

The rapid-pace growing demand for high-performance computation and big-data manipulation entails substantial increase in global power consumption, and challenging thermal management. Thus, there is a need in allocating competitive alternatives for complementary metal-oxide-semiconductor (CMOS) technologies. Superconducting platforms, such as rapid single flux quantum (RSFQ) lack electric resistance and excel in power efficiency and time performance. However, traditional RSFQs require 3D geometry for their Josephson junctions (JJs) imposing a large footprint, and hence preventing device miniaturization and increasing processing time. Here, we demonstrate that RSFQ logic circuits of planar geometry with weak-link bridges are scalable, relatively easy to process and are CMOS-compatible on a Si chip. Universal logic gates, as well as combinational arithmetic circuiting that are based on these devices are demonstrated. The power consumption and processing time of these logic circuits were as low as 0.8 nW and 13 ps, an order of magnitude improvement with respect to the equivalent traditional-RSFQ logic circuits and two orders of magnitude with respect to CMOS. The competitive performance of planar RSFQ logic circuits renders them for promising CMOS substitutes, especially in the supercomputational realm.

**KEYWORDS: Single Flux Quantum, Planar Josephson Junctions, Beyond CMOS, Data Science, Supercomputer**

Silicon-based complementary metal-oxide semiconductor (CMOS)[1] is currently the leading data processing technology. However, CMOS attractiveness is challenged with the rapidly growing requirement for high-performance computation and big-data manipulation due to the involved increase in global power consumption and thermal management.[2,3] The absence of dc electric resistance and low ac power consumption, renders superconductors as a potential CMOS substitute. Specifically, conventional rapid single flux quantum (RSFQ)[4,5] is a superconductive platform for digital logics with up to ~100 GHz[9] clocking and $10^{-18}$ J power consumption per a sum-operator device (at 5 GHz).[6,10–12] This is in comparison to ~10 GHz maximum frequency[13] and $10^{-16}$ J device average (at 5 GHz) for conventional CMOS[1,10] and $10^{-18}$ J for an advanced (but hard-to-fabricate) FinFET CMOS.[14] While CMOS operates at room temperature, RSFQ requires cryogenic temperatures. Hence, RSFQ is less suitable for personal computers, but is promising for, *e.g.*, the large market of supercomputers, which aims at the growing demand for big-data manipulation and artificial intelligence.[3] Cooling down the system entails energy loss, so a larger energy gain is required for an appealing shift from CMOS to RSFQ.

The data in RSFQ are stored as flux quanta and is manipulated with voltage and magnetic fields. RSFQ basic building blocks are Josephson junctions (JJs), in which electrons tunnel through a thin insulating medium that is sandwiched between two superconductive electrodes.

Recent RSFQ studies demonstrated energy-efficient circuits,[11,12] zero static-power,[15] circuit scalability,[16] robust chip prototypes,[17] novel approaches to JJs,[18] and attempts to increase the operational temperature.[19–22] However, JJs require a complex fabrication process and result in a relatively large footprint devices, hindering device scalability for advanced logic circuits.[16] Moreover, the complex fabrication process hinders device integration with the widespread silicon technology. Recent works propose significant processing simplification by replacing the conventional 3D JJ structures with planar weak links,[23] *e.g.*, in dc superconducting quantum interference devices.[24–26] For instance, planar high-$T_c$ superconducting bi-crystal weak links were utilized for RSFQ systems.[21,27] These studies showed that the superconducting-circuit simulator, JSIM[28] models planar devices rather precisely. However, the feasibility of planar JJs has not been fully investigated yet.[29] Thus, a rigorous demonstration of planar RSFQ logic circuits, evaluation of the power consumption and a comparison to 3D RSFQ are needed.

Here, common 3D JJs are replaced with planar weak links, reducing the RSFQ power consumption and device footprint, simplifying device fabrication, enabling Si-chip integration, and accelerating processing time. Weak links of NbN and $\alpha Mo_{1-x}Si_x$ were fabricated and characterized. The experimental parameters were used in advanced device modelling to demonstrate that substituting the common 3D Al-Al$_x$O$_y$-Al JJs[30] with planar weak links[25] reduces the RSFQ power consumption to $10^{-20}$ J for a sum operator at 5 GHz, four orders of magnitude less than conventional CMOS, while maintaining the attractive 100 GHz operational frequency. It is also illustrated that a planar integrated system of a universal half-adder logic circuit consumes two orders of magnitude less power than a conventional 3D JJ half adder. Our results indicate that reduction in power consumption is possible with suitable material selection, where materials with lower critical current ($I_c$) consume less power, while we did not find a similar dependence on the normal resistance.

RSFQ circuits are primarily composed of JJs and inductors. The single JJ Hamiltonian is: $\hat{H} = -E_J \cos\hat{\phi}$[31], where $\hat{\phi} = \hat{\phi}_1 - \hat{\phi}_2$ is the phase operator, defining the phase drop across the JJ. While the standard current-phase relationship $I = I_c \sin\phi$ holds reasonably also for weak links, we used the proper weak-link current phase relation.[31,32] An alternative approach to analyse the device behaviour is by looking at Kirchhoff's law of the equivalent circuit (Figure 1A):

$$I = I_c \sin\phi + \frac{1}{R_n}\frac{\hbar}{2e}\frac{d\phi}{dt} + C\frac{\hbar}{2e}\frac{d^2\phi}{dt^2} \quad (1)$$

Here, $I$, $I_c$, $C$, $R_n$, $\phi$, $e$, $\hbar$ are the total current through the JJ, device critical current, equivalent capacitor, and normal resistance, scalar representation of $\hat{\phi}$, electron charge and the reduced Planck's constant, respectively. Given that voltage drops ($V$) across the device are simpler to measure than phase drops, it is convenient to use: $V = \frac{\hbar}{2e}\frac{d\phi}{dt}$. Figure 1B shows a typical 3D geometry of a common JJ in RSFQ.[30] In the current work, this 3D structure was substituted by a planar weak link (Fig. 1C), where a miniaturized bridge replaced the barrier tunnel junction.[25] Figure 1D demonstrates experimental realization of such a thin (<10 nm) planar weak link NbN structure on a SiO$_2$/Si substrate, which was made through a one-step lithography process.

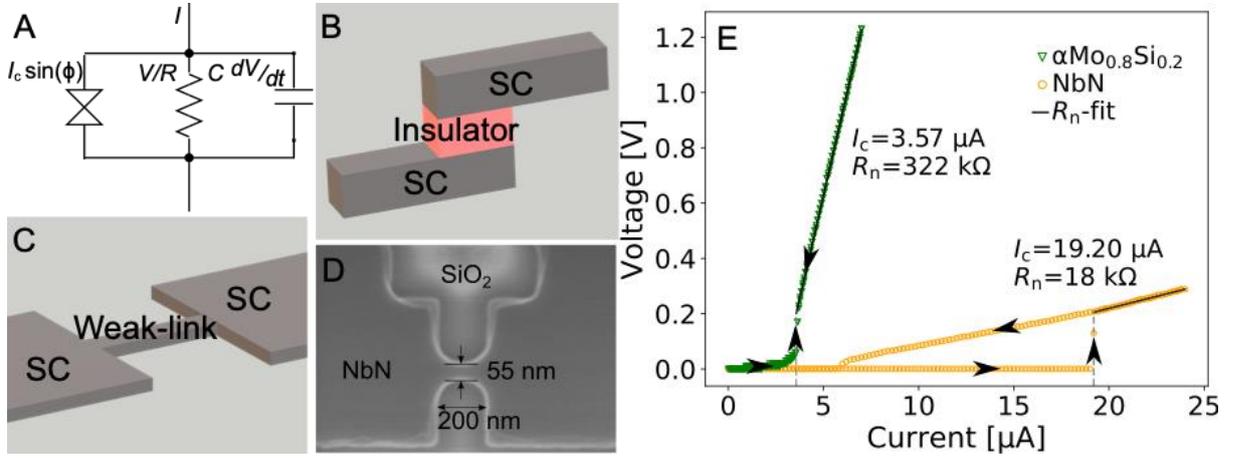

**Fig. 1| Planar Weak Links**. (**A**) JJ equivalent circuit.[31,32] (**B**) Schematic layout of a 3D JJ. (**C**) Schematic layout of a planar JJ weak-link. (**D**) Electron micrograph of a planar NbN weak link that was fabricated in one-step electron-beam lithography on a SiO$_2$/Si chip. (**E**) Experimentally measured *I-V* characteristics of planar αMo$_{0.8}$Si$_{0.2}$ and NbN weak links, from which the respective critical currents (3.57 µA and 19.20 µA) and normal resistance (322 kΩ and 18 kΩ) were extracted.

The information in RSFQ circuits is represented as a sudden $2\pi$ phase change, which is equivalent to a quantum flux change: $\Phi_0 = \frac{h}{2e}$,[33] giving rise to a measurable voltage pulse. A logic state '1' is represented by an SFQ pulse, while a '0' state is the absence of such a pulse. Commonly, the duration of an SFQ pulse can be as fast as 2 ps with 1-mV-voltage amplitude.[6] The circuit is constantly current-biased at the superconductive 0-voltage state, close to $I_c$. Thus, as the voltage pulse progresses in the circuit, it switches the JJs on and off one by one, along the pulse route.

The power consumption in RSFQ is the time integration of the voltage multiplied by the current flow across the device during a pulse firing. Intuitively, lower power consumption is obtained for lower normal resistance, and lower operation current, which is derived from $I_c$ and $R_n$, both are material and geometry dependent. Figure 1E shows experimental *I-V* curves of planar weak links for both NbN and αMo$_{0.8}$Si$_{0.2}$, from which critical current and normal resistance were extracted. The obtained values were then compared to that of Al superconductor-insulator-superconductor (SIS) junctions[30] with $I_c = 100$ µA and $R_n = 10\Omega$,[34] both indicate higher power consumption than the planar devices (Fig. 1E). Note that there has been a recent effort to reduce $I_c$ (and $R_n$) for 3D JJs.[35] However, processing such devices is significantly eve more complex than 3D JJs, while the reduced $I_cR_n$ deteriorates the time performance.

The total power consumption during logic operations is conventionally split into dynamic (directly related to computations) and static (constant circuit feed wastes). In RSFQ, dynamic power consumed during a single logic operation is theoretically approximated by the number of JJs in a circuit ($n_{jj}$), their critical current $I_c$, and the computation frequency, $f$:[36]

$$P_d^t = n_{jj} f \Phi_0 I_c. \qquad (2)$$

Likewise, theoretical RSFQ static power depends on the number of current-bias feeding lines $n_b$ (*i.e.*, $n_b = 1$ in Fig. 2):

$$P_s^t = n_b V_b I_b \qquad (3)$$

where the bias current is set here close to the critical value $I_b = 0.7 I_c$.[36]

Another dissipating power is heat waste on the shunting resistors (which are parallel to JJs, see Fig. 2). However, this dissipative power is typically a few orders of magnitude lower than the dynamic and static power consumption (see SI Section B for detail).

A simple case study for 3D and planar geometries power consumption comparison is examining the elements that interconnect neighbouring logic cells. Josephson transmission lines (JTLs) are used as pulse conductors, which introduce a particular time-delay,[37] facilitating the passage of an input SFQ pulse in either direction and potentially enhancing its sharpness. In practical applications, the JTL primary function is interconnecting cells immediately next to each other, while providing insulation between them. Figure 2A shows schematics of the JTL circuit element, where Fig. 2B illustrates its fabrication design, which includes one-step lithography for the entire circuit.

Note that RSFQ requires overdamped JJs, in which the switching current is identical for both switching from and to the superconductive states, allowing for fast recovery time after each pulse, and hence operating the device at higher frequencies.[38,39] Figure 1E shows a difference between the switching current (superconducting-to-normal transition) and retrapping current (normal-to-superconducting transition), indicating on an underdamped behaviour. This difference is much smaller for the αMo$_{0.8}$Si$_{0.2}$. Nevertheless, even the NbN JJ can behave effectively as an overdamped junction, simply by adjusting the parallel shunt resistor ($R_s$). Shunt resistance is introduced conveniently with the planar geometry via an additional lithography step of placing a metallic resistor above the junction, which is a much of a bigger challenge in the 3D JJ geometry (see Fig. 2). Details about the circuit parameters that were used to assure overdamped behaviour are given in the SI.

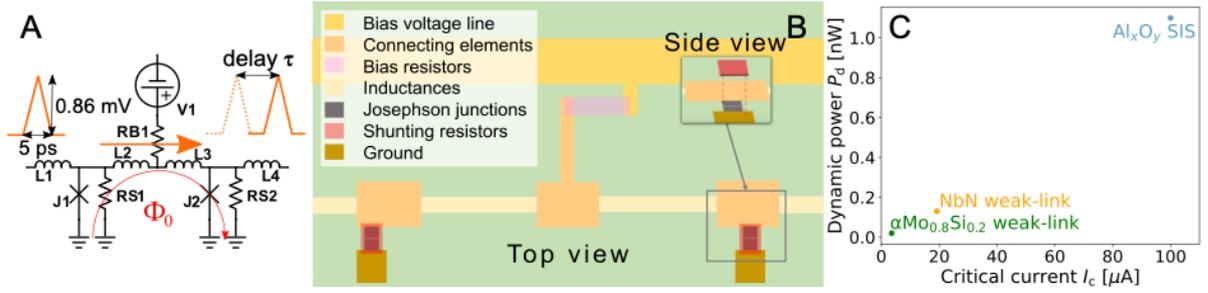

**Fig. 2 |Planar JTL.** (**A**) RSFQ transmission line (schematics). A short pulse travelling across the transmission line represents a logic bit at state '1', while its absence represents '0.' Pulses travel through the line left-to-right (orange arrow) switching JJs one by one via single flux-quantum circulation (red curl). Upon reaching the end of the JTL, some delay time gained by the travelling pulse, while the pulse shape is preserved. (**B**) Weak-link-based layout of the transmission line from frame (A) obtained in a one-step lithography process. Shunt resistance is introduced for underdamped weak links. (**C**) Simulating a JTL with $n_{jj}$ = 2 of 3D (Al) SIS and planar (NbN and $\alpha Mo_{0.8}Si_{0.2}$) weak-links allowed direct comparison between the different JTL performances. An order of magnitude difference in dynamic power and a factor of 2-3 in time delay are illustrated (see SI Section C for other power parameters). $\tau$ = 12.0, 7.0, 7.0 ps JTL delay times for Al SIS, and NbN and $\alpha Mo_{0.8}Si_{0.2}$ weak-links, correspondingly, for respective $R_s$ = 70 Ω, 350 Ω and 7 kΩ were extracted.

The device processing is detailed elsewhere[24,40-42] (see also SI Section A). The JTL power consumption that comprises conventional 3D Al-Al$_x$O$_y$-Al JJs was simulated with JoSIM,[36] with $I_c \sim$ 100 μA and $R_n \sim$ 10 Ω[43] device parameters (see SI Section C). Similar calculations were done for $\alpha Mo_{0.8}Si_{0.2}$ and NbN, based on the experimental parameters (Fig. 1). Figure 2C shows that the planar JTL dynamic power consumption is about an order of magnitude lower for a planar NbN JTL than the 3D JTL, while the $\alpha Mo_{0.8}Si_{0.2}$ power consumption is even an additional order of magnitude lower. These calculated dynamic power-consumption values agree with Equation (2), emphasizing that $P_d$ depends linearly on the critical current, which is lowest for the $\alpha Mo_{0.8}Si_{0.2}$.

NAND and NOR are universal digital logic gates (Fig. 3A). That is, these gates provide a foundation upon which all other logic functions can be built, including AND, OR, NOT, and various combinations thereof. The general purpose of universal NAND and NOR gates lies in their ability to simplify the digital-circuits design and manufacturing. Instead of producing different gate types, manufacturers can focus on a single gate type (NAND or NOR) and use it to build complex logic functions, lowering costs and increasing production efficiency.

Table I summarizes power and timing performance of planar RSFQ NAND and NOR logic gates with CMOS. To simulate device performance of similar universal gates for 'traditional' 3D Al SIS and planar NbN and $\alpha Mo_{0.8}Si_{0.2}$ geometries, the JoSIM library[36] was adopted (see SI Section D). The JoSIM library does not contain these logic gates, so the basic components were used for creating the universal gate circuits. Figure 3B shows the voltage-pulse time evolution of a two-inputs (IN1 and IN2) and clock (CLK) for successful NAND and NOR operations of the three systems (Al-SIS and planar NbN and $\alpha Mo_{0.8}Si_{0.2}$). The voltage dynamics were then used to calculate the NAND (NOR) circuit dynamic power consumption, which was 26 (28) nW, 4 (4) nW and 2 (2) nW for the three material geometries, respectively, in comparison to ≳ 1 μW of CMOS gates at the same frequency.[1,10]

Fluxon dynamics was then evaluated by careful examination of the phase dynamics (Fig. 3C), as the change in the fluxon state is accompanied by a 2π quantized phase change. Both NAND and NOR results show consistently that the αMo$_{0.8}$Si$_{0.2}$ planar circuits were the fastest (11 and 15 ps) and the NbN planar circuits were slightly slower (17 and 23 ps). The 3D Al SIS circuits were much slower (37 and 41 ps), though yet much faster than conventional Si-based CMOS gates (500 ns[44]) and comparable to the complex and slow (~2 GHz) FinFET CMOS (1.2 ps).[14] The overall phase dynamics of the two universal gates are given in Fig. 3D.

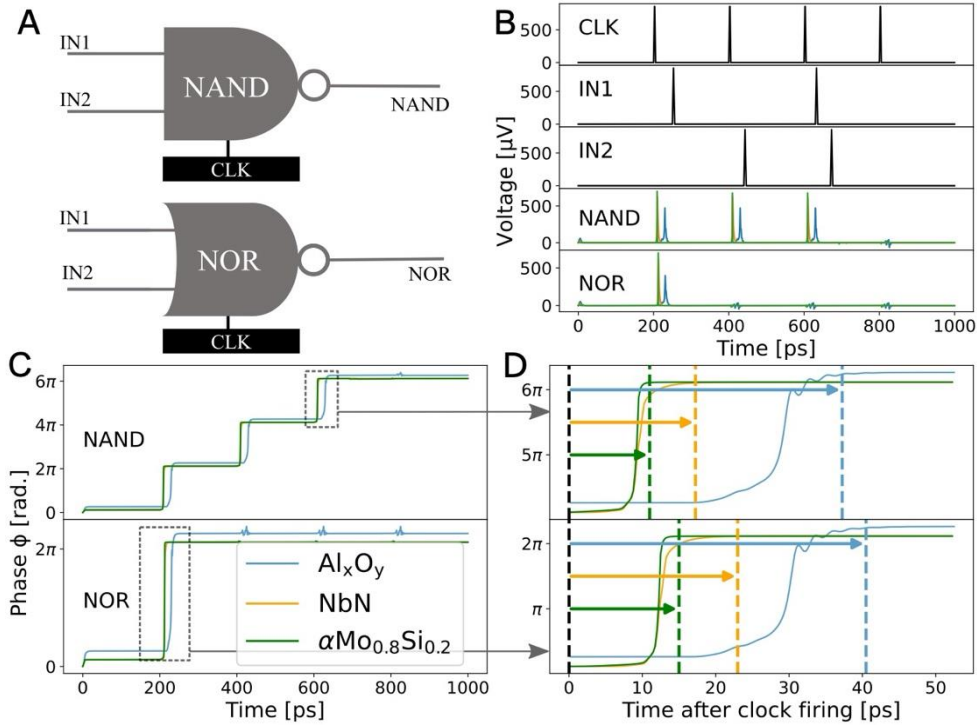

**Fig. 3 | Universal logic gates.** (**A**) Schematics of NAND and NOR logic blocks. Each gate has a clock-synchronized (CLK) dual input signals (IN1 and IN2). (**B**) Comparison of simulated voltage-pulse dynamics in NAND and NOR gates of different JJ materials and geometries: 3D alumina JJ (blue), and planar NbN (orange) and αMo$_{0.8}$Si$_{0.2}$ (green) JJs, showing successful logic operation in all cases. Clock and input signals appear in black. The planar devices demonstrate advantageous time response and power consumption, with $P_d$ = 26, 4, 2 nW (NAND) and $P_d$ = 28, 4, 2 nW (NOR) for 3D Al SIS and NbN, and αMo$_{0.8}$Si$_{0.2}$ planar weak links, respectively (in comparison to $P_d^{CMOS} \gtrsim$ 10 μW for conventional CMOS[38] and $P_d^{FinFET} \gtrsim$ 1 μW for the complex FinFET CMOS technology[45]). CMOS-relevant 5-GHz operational frequency was set. The power consumption is proportional to $I_c$ in agreement with Eq. (2). (**C**) Simulated phase propagation showed consistent behaviour. (**D**) A closer look at the phase evolution (C) demonstrated characteristic delay time of τ = 37, 17, 11 ps (NAND) and τ = 41, 23, 15 ps (NOR) for 3D Al SIS and NbN, and αMo$_{0.8}$Si$_{0.2}$ planar weak links, respectively, in comparison to 500 ps and 1.2 ps of conventional CMOS[44] and FinFET CMOS technology[14], respectively. CMOS-relevant 5-GHz operational frequency and 4.2 K were used (JoSIM[36]).

Complementary to universal gates, adding numbers is a significant capability of logic circuits. Half adders are basic combinational arithmetic circuits that handle multiple input bits and carry inputs from previous stages. As opposed to NAND and NOR, half-adders have a dual-signal output: sum and carry, which are the output signals of the XOR and AND components, respectively (Fig. 4A, see SI Section D for details). Figure 4B shows a pulse-signal voltage evolution in a half-adder, which operates as expected. The voltage dynamics allows to extract the dynamic power, which was 37 nW for 3D Al SIS, 8 nW for NbN planar circuit and only 1 nW for a αMo$_{0.8}$Si$_{0.2}$ planar circuit. These values are at the order of (Al SIS) or much smaller than (planar weak-links) the $\gtrsim$1 μW CMOS half-adder dynamic power.[1,10]

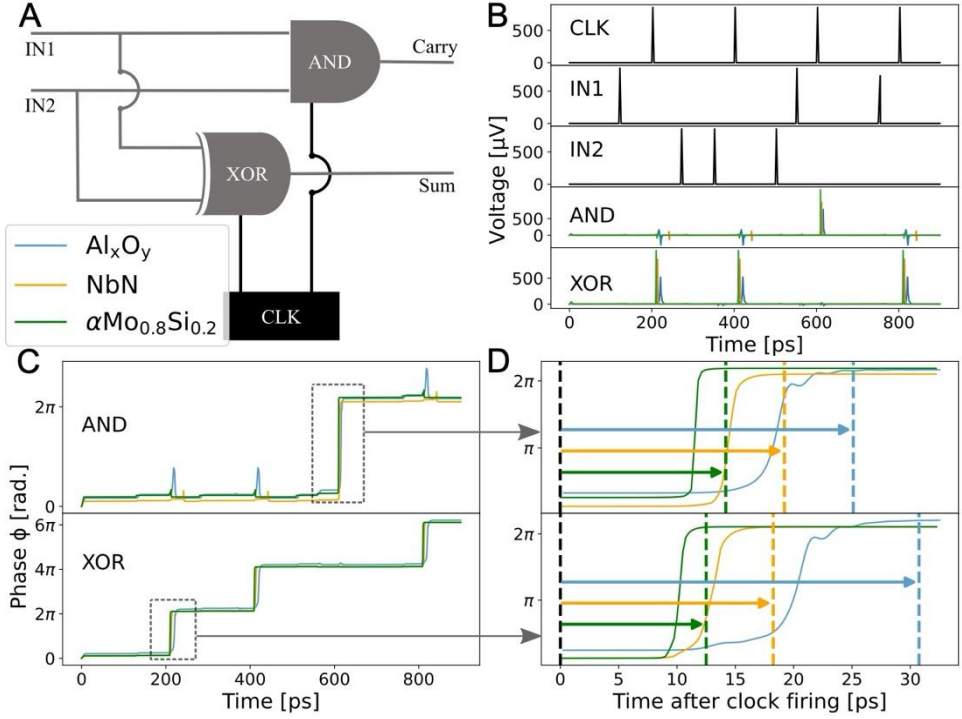

**Fig. 4 | Half-adder.** (**A**) Schematics of a half-adder logic block. A dual input signal is fed into synchronized (CLK) AND and XOR gates (black), providing Carry (AND) and Sum (XOR) signals. (**B**) Simulated voltage-pulse dynamics in half-adder gates of different JJ materials and geometries: 3D alumina JJ (blue), and planar NbN (orange) and $\alpha Mo_{0.8}Si_{0.2}$ (green) JJs, showing successful logic operation in all cases. The planar devices demonstrate faster response and lower power consumption than the 3D geometry, with $P_d$ = 37, 8, 1 nW for 3D Al SIS and NbN, and $\alpha Mo_{0.8}Si_{0.2}$ planar weak links, respectively (in comparison to $P_d^{CMOS} \gtrsim 10$ µW for conventional CMOS[46] and $P_d^{FinFET} \gtrsim 1$ µW for complex FinFET CMOS technology).[45] CMOS-relevant 5-GHz operational frequency was set. (**C**) Simulated phase propagation showed consistent behaviour. (**D**) A closer look at the phase evolution (C) demonstrated characteristic delay time of τ = 31, 18, 13 ps for 3D Al SIS and NbN, and $\alpha Mo_{0.8}Si_{0.2}$ planar weak links, respectively, in comparison to 500 ps and 30 ps of conventional CMOS[40] and FinFET CMOS technology,[47] respectively. CMOS-relevant 5-GHz operational frequency and 4.2 K were used (JoSIM[36]).

Figure 4C allows a careful look at the phase dynamics in the three RSFQ systems. Here, the time delay between the CLK and the output signal was 31 ps (3D Al SIS), 18 ps (planar NbN) and as fast as 13 ps for $\alpha Mo_{0.8}Si_{0.2}$ planar circuits. These devices were much faster than the conventional Si-based CMOS gates (500 ns[44]) and comparable to the advanced 3D FinFET CMOS (1.2 ps[14]), which allows > 10 GHz operational frequency. Table 1 presents a comprehensive comparison between the different device performance parameters. Note that the maximal operation frequency ($f_0$) can be estimated from the system time delay, which is the accumulated contribution of the JJs along the pulse propagation. For each RSFQ logic circuit element, the maximal operational frequency is dictated by the Josephson frequency $f_{max} = I_c R_n / \Phi_0$[48]. Thus, planar junctions offer a competitive combination of low power consumption and fast calculation times.

Despite the similarity between the theoretically predicted and simulated power consumptions, there are some discrepancies that are introduced because the theory assumes an ideal JJ and perfect voltage and current pulse shapes, so that:

$$\Phi_0 = \int V(t)\, dt \tag{4}$$

However, the circuit resistors redistribute the simulated voltage/current, so that $P_d = \sum_i \int dt\, V_i(t) I_i(t)$ does not match $P_d^t = n_{jj} f \Phi_0 I_c$.

|  | Geometry | 3D | Planar | | Planar | 3D |
|---|---|---|---|---|---|---|
|  | Material/Platform | Al | NbN | $\alpha Mo_{0.8}Si_{0.2}$ | CMOS[44,46] | FinFET CMOS[14,45,47] |
|  | $I_c$, μA | 100.0 | 19.2 | 3.6 |  |  |
| NAND | $P_d$, nW | 25.9 | 3.8 | 2.1 | $63 \cdot 10^3$ | $1 \cdot 10^3$ |
|  | $P_d^t$, nW | 24.8 | 4.8 | 0.9 |  |  |
|  | $P_s$, nW | 6752.0 | 1269.6 | 754.5 |  |  |
|  | $P_s^t$, nW | 2184.0 | 419.3 | 78.0 |  |  |
|  | $P_{sh}$, pW | 12.6 | 2.6 | 1.9 |  |  |
|  | τ, ps | 37.3 | 17.3 | 11.0 | 583.0 | 2.1 |
| NOR | $P_d$, nW | 28.0 | 4.3 | 1.7 | $63 \cdot 10^3$ | $1 \cdot 10^3$ |
|  | $P_d^t$, nW | 21.7 | 4.2 | 0.8 |  |  |
|  | $P_s$, nW | 6667.0 | 1254.1 | 466.2 |  |  |
|  | $P_s^t$, nW | 2002.0 | 384.4 | 71.5 |  |  |
|  | $P_{sh}$, pW | 14.5 | 1.8 | 1.3 |  |  |
|  | τ, ps | 40.5 | 23.0 | 15.0 | 551.0 | 1.0 |
| Half-adder | $P_d$, nW | 36.6 | 7.5 | 0.8 | $10 \cdot 10^3$ | $1 \cdot 10^3$ |
|  | $P_d^t$, nW | 45.5 | 8.7 | 1.6 |  |  |
|  | $P_s$, nW | 7020.0 | 1254.0 | 233.4 |  |  |
|  | $P_s^t$, nW | 5096.0 | 978.4 | 181.9 |  |  |
|  | $P_{sh}$, pW | 14.4 | 2.6 | 0.3 |  |  |
|  | τ, ps | 30.8 | 18.3 | 12.5 | 420.0 | 30.0 |

**Table 1 | Competitive planar RSFQ performance.** Comparison of dynamic ($P_d$), static ($P_s$), and shunt ($P_{sh}$) power consumption/release and time delay (τ) of RSFQ circuits based on 3D Al-based SIS, planar NbN, and planar $\alpha Mo_{0.8}Si_{0.2}$ JJs. The circuits were simulated in JoSIM.[36] Additionally, dynamic and static powers were estimated from Equations (**2**) and (**3**) (denoted with □$^t$). We also provide equivalent experimental values for circuits implemented on two CMOS platforms: conventional planar CMOS and advanced 3D FinFET CMOS.

To conclude, RSFQ circuits of planar JJ geometry were proved to allow successful logic-data manipulation. The power consumption of combinatorial binary arithmetic circuits and universal logic gates was found to be up to two orders of magnitude lower than similar circuits that are made of conventional 3D Al-based SIS JJs and up to four orders of magnitude lower than equivalent CMOS circuits. A similar trend was observed in the time delay of these gates. Therefore, these results demonstrate that planar JJs circuiting, which is simpler for fabrication than 3D geometries, Si-compatible, and suitable for device miniaturization/circuiting scaling up are competitive with respect to both operational frequency and power consumption and are therefore promising, *e.g.*, for the growing realm of supercomputers.

See Supplementary Information (SI) for the device fabrication and simulation details.

N.G. and R.M. acknowledge the fellowship support from Hellen Diller Quantum Center and Department of Materials Science & Engineering. We would like to thank the following personnel for technical support and fruitful discussions: Dr. Guy Ankonina (sputtering and ellipsometry); Dr. Adi Goldner and Dr. Roni Winik (device fabrication). Y.I. acknowledges support from the Zuckerman STEM Leadership Program, Grand Technion Energy Program, Technion's Hellen Diller Quantum Center and Pazy Research Foundation Grant No. 149-2020.

## AUTHOR DECLARATION

### Conflict of Interest
The authors have no conflict to discuss

### Author Contributions
**Nikolay Gusarov:** Simulation and writing manuscript. **Rajesh Mandal:** Guidance in simulation and analysis, project management. **Issa Salameh:** Guidance in simulation. **Itamar Holzman**: Device fabrication and measurement. **Shahar Kvatinsky**: Supervision of simulation. **Yachin Ivry:** Conceptualization, supervision and writing manuscript.

## DATA AVAILABILITY

Any supporting data and additional information could be directly requested to the corresponding author.

## REFERENCES

[1] H. H. Radamson, H. Zhu, Z. Wu, X. He, H. Lin, J. Liu, J. Xiang, Z. Kong, W. Xiong, J. Li et al., "State of the art and future perspectives in advanced cmos technology," Nanomaterials, vol. 10, no. 8, p. 1555, (2020).

[2] NVIDIA, "NVIDIA A100 Tensor Core GPU architecture [white paper]," 2020, last accessed 29 May 2023. [Online]. Available: https://images.nvidia.com/aem-dam/en-zz/Solutions/data-center/nvidia-ampere-architecture-whitepaper.pdf

[3] F. H. Khan, M. A. Pasha, and S. Masud, "Advancements in microprocessor architecture for ubiquitous ai—an overview on history, evolution, and upcoming challenges in ai implementation," Micromachines, vol. 12, no. 6, p. 665, (2021).

[4] O. Mukhanov, V. Semenov, and K. Likharev, "Ultimate performance of the rsfq logic circuits," IEEE Transactions on Magnetics, vol. 23, no. 2, pp. 759–762, (1987).

[5] K. K. Likharev and V. K. Semenov, "Rsfq logic/memory family: A new josephson-junction technology for sub-terahertzclock-frequency digital systems," IEEE Transactions on Applied Superconductivity, vol. 1, no. 1, pp. 3–28, (1991).

[6] M. Pedram, "Superconductive single flux quantum logic devices and circuits: Status, challenges, and opportunities," in 2020 IEEE International Electron Devices Meeting (IEDM). IEEE, 2020, pp. 25–7.

[7] P. Barla, V. K. Joshi, and S. Bhat, "Spintronic devices: a promising alternative to cmos devices," Journal of Computational Electronics, vol. 20, no. 2, pp. 805–837, (2021).

8. D. Lyles, P. Gonzalez-Guerrero, M. G. Bautista, and G. Michelogiannakis, "Past-noc: A packet-switched superconducting temporal noc," IEEE Transactions on Applied Superconductivity, (2023).
9. W. Chen, A. Rylyakov, V. Patel, J. Lukens, and K. Likharev, "Rapid single flux quantum t-flip flop operating up to 770 ghz," IEEE Transactions on Applied Superconductivity, vol. 9, no. 2, pp. 3212–3215, (1999).
10. N. Takeuchi, T. Yamae, C. L. Ayala, H. Suzuki, and N. Yoshikawa, "Adiabatic quantum-flux-parametron: A tutorial review," IEICE Transactions on Electronics, vol. 105, no. 6, pp. 251–263, (2022).
11. O. A. Mukhanov, "Energy-efficient single flux quantum technology," IEEE Transactions on Applied Superconductivity, vol. 21, no. 3, pp. 760–769, (2011).
12. D. S. Holmes, A. L. Ripple, and M. A. Manheimer, "Energy-efficient superconducting computing—power budgets and requirements," IEEE Transactions on Applied Superconductivity, vol. 23, no. 3, pp. 1701610–1701610, (2013).
13. S. Lueangsongchai and S. Tooprakai, "Design high speed and low power hybrid full adder circuit," in 2018 18th International Symposium on Communications and Information Technologies (ISCIT). IEEE, 2018, pp. 22–25.
14. A. Lazzaz, K. Bousbahi, and M. Ghamnia, "Performance analysis of finfet based inverter, nand and nor circuits at 10 nm, 7 nm and 5 nm node technologies," Facta universitatis-series: Electronics and Energetics, vol. 36, no. 1, pp. 1–16, (2023).
15. D. Kirichenko, S. Sarwana, and A. Kirichenko, "Zero static power dissipation biasing of rsfq circuits," IEEE Transactions on Applied Superconductivity, vol. 21, no. 3, pp. 776–779, (2011).
16. S. K. Tolpygo and V. K. Semenov, "Increasing integration scale of superconductor electronics beyond one million josephson junctions," in Journal of Physics: Conference Series, vol. 1559, no. 1. IOP Publishing, 2020, p. 012002.
17. H. Hayakawa, N. Yoshikawa, S. Yorozu, and A. Fujimaki, "Superconducting digital electronics," Proceedings of the IEEE, vol. 92, no. 10, pp. 1549–1563, (2004).
18. I. Salameh, E. G. Friedman, and S. Kvatinsky, "Superconductive logic using 2$\phi$—josephson junctions with half flux quantum pulses," IEEE Transactions on Circuits and Systems II: Express Briefs, vol. 69, no. 5, pp. 2533–2537, (2022).
19. M. Y. Kupriyanov and K. K. Likharev, "Josephson effect in high-temperature superconductors and in structures based on them," Soviet Physics Uspekhi, vol. 33, no. 5, p. 340, (1990).
20. G. Blatter, M. V. Feigel'man, V. B. Geshkenbein, A. I. Larkin, and V. M. Vinokur, "Vortices in high-temperature superconductors," Reviews of modern physics, vol. 66, no. 4, p. 1125, (1994).
21. B. Oelze, B. Ruck, M. Roth, R. Dömel, M. Siegel, A. Y. Kidiyarova-Shevchenko, T. Filippov, M. Y. Kupriyanov, G. Hildebrandt, H. Töpfer et al., "Rapid single-flux-quantum balanced comparator based on high-t c bicrystal josephson junctions," Applied physics letters, vol. 68, no. 19, pp. 2732–2734, (1996).
22. R. Gross, L. Alff, A. Beck, O. Froehlich, D. Koelle, and A. Marx, "Physics and technology of high temperature superconducting josephson junctions," IEEE transactions on Applied Superconductivity, vol. 7, no. 2, pp. 2929–2935, (1997).
23. K. K. Likharev, "Superconducting weak links," Rev. Mod. Phys. 51, 101, 1979.
24. I. Holzman and Y. Ivry, "On-chip integrable planar nbn nanosquid with broad temperature and magnetic-field operation range," AIP Advances, vol. 9, no. 10, p. 105028, (2019).
25. R. Vijay, E. Levenson-Falk, D. Slichter, and I. Siddiqi, "Approaching ideal weak link behavior with three dimensional aluminum nanobridges," Applied Physics Letters, vol. 96, no. 22, (2010).
26. Z. Ivanov, P. Nilsson, D. Winkler, J. Alarco, T. Claeson, E. Stepantsov, and A. Y. Tzalenchuk, "Weak links and dc squids on artificial nonsymmetric grain boundaries in yba2cu3o7-$\delta$," Applied physics letters, vol. 59, no. 23, pp. 3030–3032, (1991).

[27] C. D. Shelly, P. See, J. Ireland, E. J. Romans, and J. M. Williams, "Weak link nanobridges as single flux quantum elements," Superconductor Science and Technology, vol. 30, no. 9, p. 095013, (2017).

[28] E. S. Fang, "Jsim superconducting circuit simulator," 2017, last accessed 23 May 2023. [Online]. Available: https://github.com/coldlogix/jsim/blob/master/README.md

[29] Collins, Jonathan A., Calum S. Rose, and Alessandro Casaburi. "Superconducting Nb Nanobridges for Reduced Footprint and Efficient Next-Generation Electronics." IEEE Transactions on Applied Superconductivity 33, no. 1 (2022): 1-8.

[30] L. Fritzsch, H. Elsner, M. Schubert, and H. Meyer, "Sns and sis josephson junctions with dimensions down to the submicron region prepared by a unified technology," Superconductor Science and Technology, vol. 12, no. 11, p. 880, (1999).

[31] A. A. Golubov, M. Yu. Kupriyanov, and E. Il'ichev, "The current-phase relation in Josephson junctions," Rev. Mod. Phys. 76, 411, 2004.

[32] Vijay, R., E. M. Levenson-Falk, D. H. Slichter, and I. Siddiqi. "Approaching ideal weak link behavior with three dimensional aluminum nanobridges." Applied Physics Letters 96, no. 22 (2010).

[33] N. K. Katam, J. Kawa, and M. Pedram, "Challenges and the status of superconducting single flux quantum technology," in 2019 Design, Automation & Test in Europe Conference & Exhibition (DATE). IEEE, 2019, pp. 1781–1787.

[34] I.I. Soloviev, S.V. Bakurskiy, V.I. Ruzhickiy, N.V. Klenov, M.Yu. Kupriyanov, A.A. Golubov, O.V. Skryabina, and V.S. Stolyarov, "Miniaturization of Josephson Junctions for Digital Superconducting Circuits," Phys. Rev. Applied, 16, 044060, 2021.

[35] Yohannes, D., M. Renzullo, J. Vivalda, A. C. Jacobs, M. Yu, J. Walter, A. F. Kirichenko, I. V. Vernik, and O. A. Mukhanov. "High density fabrication process for single flux quantum circuits." Applied Physics Letters 122, no. 21 (2023).

[36] J. A. Delport, K. Jackman, P. Le Roux, and C. J. Fourie, "Josim—superconductor spice simulator," IEEE Transactions on Applied Superconductivity, vol. 29, no. 5, pp. 1–5, (2019).

[37] T. Ortlepp and F. H. Uhlmann, "Gate delay jitter in superconductor electronics," in Information technology and electrical engineering-devices and systems, materials and technologies for the future, vol. 51. ilmedia, 2009, p. p. Session 3.5.

[38] T. Ortlepp, "General design aspects of integrated superconductor electronics," Cryogenics, vol. 49, no. 11, pp. 648–651, (2009).

[39] P. Febvre, D. Bouis, N. De Leo, M. Fretto, A. Sosso, and V. Lacquaniti, "Electrical parameters of niobium-based overdamped superconductor-normal metal-insulator-superconductor josephson junctions for digital applications," Journal of Applied Physics, vol. 107, no. 10, p. 103927, (2010).

[40] M. Suleiman, E. G. D. Torre, and Y. Ivry, "Flexible amorphous superconducting materials and quantum devices with unexpected tunability," arXiv: 2002. 10297, (2020).

[41] M. Suleiman, M. F. Sarott, M. Trassin, M. Badarne, and Y. Ivry, "Nonvolatile voltage-tunable ferroelectric-superconducting quantum interference memory devices," Applied Physics Letters, vol. 119, no. 11, p. 112601, (2021).

[42] Y. Ivry, C.-S. Kim, A. E. Dane, D. De Fazio, A. N. McCaughan, K. A. Sunter, Q. Zhao, and K. K. Berggren, "Universal scaling of the critical temperature for thin films near the superconducting-to-insulating transition," Physical Review B, vol. 90, no. 21, p. 214515, (2014).

[43] C. J. Fourie, K. Jackman, M. M. Botha, S. Razmkhah, P. Febvre, C. L. Ayala, Q. Xu, N. Yoshikawa, E. Patrick, M. Law et al., "Coldflux superconducting eda and tcad tools project: Overview and progress," IEEE Transactions on Applied Superconductivity, vol. 29, no. 5, pp. 1–7, (2019).

[44] G. Dai, W. Xie, X. Du, M. Han, T. Ni, and D. Wu, "Memristor-based d-flip-flop design and application in built-in self-test," Electronics, vol. 12, no. 14, p. 3019, (2023).

[45] R. K. Maurya and B. Bhowmick, "Review of FinFET devices and perspective on circuit design challenges," Silicon, vol. 14, no. 11, pp. 5783–5791, (2022).

[46] H. Waris, C. Wang, and W. Liu, "High-performance approximate half and full adder cells using nand logic gate," IEICE Electronics Express, vol. 16, no. 6, pp. 20190043–20190043, (2019).

[47] M. Satish and K. V. Patel, "Power reduction in finfet half adder using svl technique in 32nm technology," in 2019 4th MEC International Conference on Big Data and Smart City (ICBDSC). IEEE, 2019, pp. 1–5.

[48] Kaplunenko, V. K. "Fluxon interaction in an overdamped Josephson transmission line." Applied physics letters 66, no. 24 (1995): 3365-3367.

[49] C. J. Fourie, "Extraction of dc-biased sfq circuit verilog models," IEEE Transactions on Applied Superconductivity, vol. 28, no. 6, pp. 1–11, (2018).

[50] Banerjee, Archan, Luke J. Baker, Alastair Doye, Magnus Nord, Robert M. Heath, Kleanthis Erotokritou, David Bosworth, Zoe H. Barber, Ian MacLaren, and Robert H. Hadfield. "Characterisation of amorphous molybdenum silicide (MoSi) superconducting thin films and nanowires." Superconductor Science and Technology 30, no. 8 (2017): 084010.

# Supplementary Information

**Low-power Rapid Planar superconducting Logic Devices**

Nikolay Gusarov,[1] Rajesh Mandal,[1,2] Issa Salameh,[3] Itamar Holzman,[1,2] Shahar Kvatinsky,[3] and Yachin Ivry[1,2,4,*]

**AFFLIATIONS**

[1]Department of Materials Science and Engineering, Technion - Israel Institute of Technology, Haifa 3200003, Israel

[2]Solid State Institute, Technion – Israel Institute of Technology, Haifa 3200003, Israel

[3]Andrew and Erna Viterbi Faculty of Electrical and Computer Engineering, Technion - Israel Institute of Technology, Haifa 3200003, Israel

[4]Department of Materials Science and Engineering, University of California, Berkeley, CA 94720, USA

***Author to whom correspondence should be addressed: ivry@technion.ac.il**

## A. Fabrication and Current-Voltage measurements

Thin films of NbN (4 nm) (see Fig. 1C and D) and $\alpha Mo_{0.8}Si_{0.2}$ (30 nm) are deposited on Si wafer coated with 338 nm $SiO_2$ using ATC2200 (AJA International Inc. MA, USA) DC magnetron sputtering from pure Nb and $Mo_{0.8}Si_{0.2}$ targets (99.95%) respectively. Growth optimization of the $\delta$-NbN has been done based on our previous work[42] in a reactive gas environment (Ar : $N_2$) via the control of dc power, substrate temperature, gas pressure and deposition time. More details of the growth parameters of NbN can be found here.[24] $Mo_{0.8}Si_{0.2}$ has been grown with off-axis geometry and at room temperature to retain its amorphous nature. Details of the growth of $Mo_{0.8}Si_{0.2}$ can be found in previous reports.[40,41]

Film thickness was determined by using a five-oscillator fit (one Drude, one TaucLorentz, one Gauss-Lorentz, and two Gaussians) of varying-angle spectroscopy ellipsometry (VASE) measurements of the optical constants (M-2000 by J. A. Woollam, NE USA).

After the deposition JJs are fabricated with electron beam lithography method in clean room. PMMA 950A3 electron-beam resist was applied by spin coating, followed by 2-min post baking. A thin layer of 'e-spacer' was used to avoid charging under the electron microscope. Next, the film was patterned with electron-beam lithography 530 $\frac{\mu C}{cm^2}$ dose of 150 pA for NbN and 610 $\frac{\mu C}{cm^2}$ dose of 250 pA for $Mo_{0.8}Si_{0.2}$ layer at 100 kV, Raith EBPG 5200) and the PMMA was developed. Reactive-ion etching (RIE) with $CF_4$ gas transferred the pattern to both of the $Mo_{0.8}Si_{0.2}$ and NbN layer. To prevent degradation of the PMMA mask, the RIE process was done on a cold stage (1°C). In addition, because the etching process generates heat, we split the process to 3 x 20-sec segments, which allowed the PMMA to cool down. The PMMA was then removed using acetone.[24, 40, 41]

High resolution scanning electron Microscopy (SEM) is used to take the image and lateral measurements of the JJs (Hitachi S-4800 or Zeiss Ultra Plus).

JJ characteristics and *I-V* measurement (see Figure 1E) were performed in BF-LD 250 dilution refrigerator (BlueFors Cryogenics, Finland) at the temperature of 5K for NbN and 3K for αMo$_{0.8}$Si$_{0.2}$. Despite the temperature differing from the 4.2K (the temperature in the simulations), we expect the experimental data to yield the correct values with the discrepancy acceptable for the current work[40, 50]. Electric biasing was done with MFLI Lock-in Amplifier (Zurich Instruments, Switzerland).

**B. Calculations description**

The delay time $\tau$ is the difference between the start of the clock's pulse and the time when the last JJ (along the pulse's route) shifts its phase by $2\pi$ (with $10^{-2}$ threshold precision).

The simulated static powers (without superscript "t") are calculated via integrating product of corresponding voltage and current recorded in the simulations and then averaging it over the simulation time-span. Integration time-step is 0.25 ps. Thereby, dynamic power is obtained via integrating the product of the voltages and currents on all of the JJs. Static power corresponds to the heat dissipation on resistors due to the current biasing $I_b$. Shunting power represents dissipation on shunting resistors (parallel to JJs).

Theoretical estimations for the powers (with superscript "t") are obtained with Equations (**2**) and (**3**) with all the variables knows for each circuit (critical current $I_c$, static biasing current $I_b = 0.7I_c$, corresponding bias voltage $V_b$, operational frequency $f$, the number of JJs $n_{jj}$, and the number of bias resistors $n_b$).

**C. JTL**

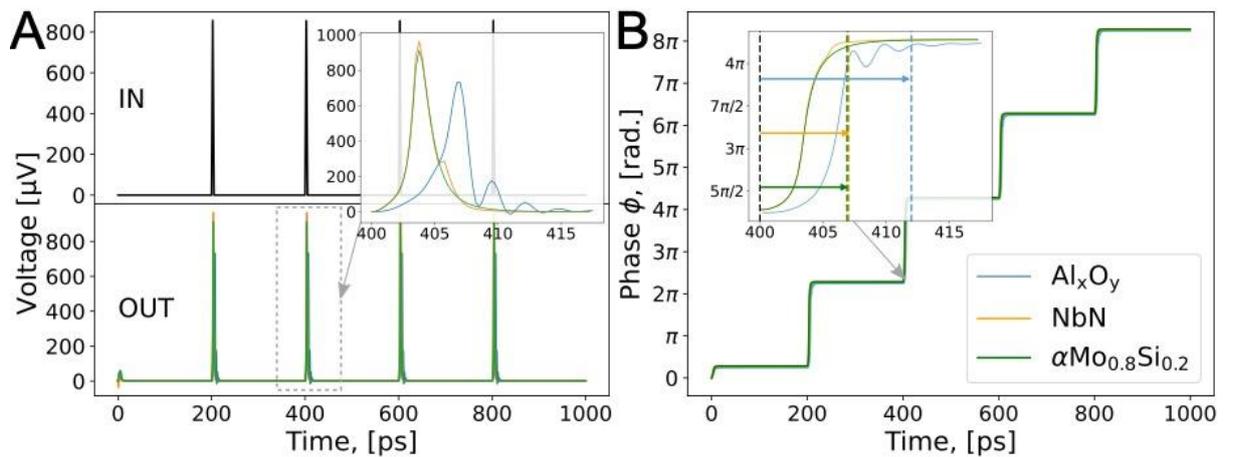

**Fig. SI 1| JTL interconnecting element.** (**A**) Pulses through JTL in simulations with different JJ models. (**B**) Phases of the last JJ in JTL in simulations with different JJ models.

Transmission line is the most basic RSFQ gate. Therefore, it demonstrates all the differences in the platforms' properties straightforwardly. Specifically, one may clearly see the factors of the difference between the powers and the delay times (see Table SI1).

In Fig. SI1 A and B, one may clearly see the distinctive oscillatory behaviour of the Al SIS platform. Such behaviour is preserved independently of the shunting resistance $R_s$ which is exactly partially used to reduce it[49]. This is observed due to the internal parameters of the "classic" Al-JJ taken from previous report.[38] Note, that it does not play an important role for the eventual powers and the delay time.

|  | Al SIS | NbN weak-link | $\alpha Mo_{0.8}Si_{0.2}$ weak-link |
|---|---|---|---|
| $P_d$, nW | 2.64 | 0.60 | 0.02 |
| $P_d^t$, nW | 2.07 | 0.40 | 0.07 |
| $P_s$, nW | 1126 | 87 | 16 |
| $P_s^t$, nW | 364 | 69 | 12 |
| $P_{sh}$, pW | 2566.1 | 3.0 | 1.4 |
| $\tau$, ps | 12.0 | 7.0 | 7.0 |

**Table SI1|** JTL, number of JJs and shunts is 2; number of bias resistors 1, frequency 5 GHz

**D. Circuit Simulations in JoSIM**

The simulations were conducted in JoSIM circuit simulator together with Cold Flux Logic Cell Library for MIT-LL SFQ Process (as a part of IAPRA Super Tools Deliverable). Each computational circuit tested in the current paper was designed as a standard RSFQ cell allowing for straightforward interconnection with the other cells and scale-up. Discussed in the main text JTL line was used to interconnect single gates where necessary. Each input source was connected the way it created potential between the common ground and the corresponding port. The input source then was subsequently attached to the source-cell stabilizing the input pulses and the load-cell providing necessary load before the main computational circuit. To correctly guide the signal from the circuit, load-cells were attached to the outputs as well which, subsequently, led to sink-cells leaking to the ground.

Internal JoSIM compiler simulated circuits at 1000 ps-time domain with a 0.025ps time-step. Such resolution allowed for a smooth analysis of 5ps-long RSFQ pulses behaving as continuous. The external noise level corresponded to the thermal noise at the 5K temperature. Altogether, our simulations, behaving in both all the extremal and normal settings as expected, are considered to be a sufficient approximation of the actual RSFQ-circuits response.